\documentclass[aps,pra,reprint,superscriptaddress]{revtex4-2}
\usepackage{amsmath,amssymb,bm}
\usepackage{dsfont}  % \mathds{1}: double-struck one (identity operator)
\usepackage{graphicx}
\graphicspath{{figures/}}
\usepackage{xcolor}
\usepackage[colorlinks=true,allcolors=blue]{hyperref}
\usepackage{orcidlink}
\newcommand{\Sx}{S^{x}}
\newcommand{\Sy}{S^{y}}
\newcommand{\Sz}{S^{z}}
\newcommand{\hb}{\bar h}
\newcommand{\qb}{\bar q}
\begin{document}

\title{A two-number rule for spin squeezing in imperfect atom arrays}

\author{Hoang-Long Nguyen\,\orcidlink{0009-0005-7922-9266}}
\affiliation{Identity Quantum Computing JSC, 117 Tran Duy Hung Street, Hanoi 10000, Vietnam}
\author{Tuan-Vu Truong\,\orcidlink{0009-0005-6357-6060}}
\affiliation{Identity Quantum Computing JSC, 117 Tran Duy Hung Street, Hanoi 10000, Vietnam}
\affiliation{College of Natural Sciences, Can Tho University, Can Tho, Vietnam}
\date{\today}

\begin{abstract}
Arrays of neutral atoms with Ising interactions, generated for example by Rydberg dressing, can produce spin-squeezed states for quantum-enhanced sensing. Real arrays, however, contain holes and edges and suffer from decoherence, and no closed-form result predicts how much squeezing a given array can reach. Using the exact solution of Ising dynamics from a product state, we show that the optimal squeezing is set by two statistics of the couplings, the effective number of partners of each atom, $\kappa$, and the spread of the local fields, $\epsilon$. A formula $F(\kappa,\epsilon)$ without fitted parameters predicts the optimal squeezing and the optimal time for two-dimensional arrays with couplings that decay with distance, including arrays with holes and arrays of different shapes. We find that holes reduce the squeezing by more than the loss of atoms alone, and we propose two design rules. A given number of atoms should be arranged as compactly and with as high a coordination as possible, and the holes should be filled by rearranging the atoms, even at the price of losing some of them. The formula extends to dephasing and spontaneous emission, and it explains how the soft-core radius of Rydberg-dressed couplings sets the squeezing, reproducing the coherent prediction for the arrays of a recent experiment.
\end{abstract}
\maketitle

% =====================================================================================
\section{Introduction}
Spin squeezing allows atomic sensors to surpass the standard quantum limit, and the amount of squeezing is quantified by the Wineland parameter~\cite{Wineland1994}. In the simplest picture, one-axis twisting, every atom interacts equally with every other atom and the squeezing grows with the number of atoms~\cite{KitagawaUeda1993}. Arrays of neutral atoms offer a natural platform for this, with interactions that decay with distance. Ising interactions generated by Rydberg dressing were proposed for squeezing in optical lattice clocks~\cite{Gil2014} and have produced squeezing in arrays of atomic ensembles~\cite{Hines2023} and in arrays of single atoms~\cite{Eckner2023}. Dipolar exchange interactions also generate squeezing in arrays of Rydberg atoms~\cite{Bornet2023}, and they were predicted to do so in gases of polar molecules~\cite{Bilitewski2021}.

For Ising couplings, the dynamics from a product state can be solved exactly for any coupling matrix~\cite{FossFeig2013,vandenWorm2013}. Building on this solution, Foss-Feig \textit{et al.} showed that couplings of finite range still squeeze, and identified the spread of the local fields as a factor that limits the squeezing~\cite{FossFeig2016}. Foss-Feig \textit{et al.} also showed that for Ising couplings that decay faster than $1/r^D$ in $D$ dimensions, the squeezing saturates as the number of atoms grows~\cite{FossFeig2016}, whereas Perlin \textit{et al.} showed that adding spin-exchange couplings lets the squeezing keep growing~\cite{Perlin2020}. Young \textit{et al.} showed that Rydberg-dressed couplings squeeze more as more atoms fall within the soft-core radius~\cite{Young2023}. These works establish which couplings squeeze, but they do not give a closed-form prediction of the optimal squeezing of a given array, and they leave open how the arrangement of the atoms sets it.

This question matters because real arrays are imperfect. Stochastic loading leaves holes~\cite{Schlosser2001}, and rearranging the atoms into a defect-free array takes time and loses atoms~\cite{Barredo2016,Endres2016}. Random filling has been studied in one dimension with empirical fits, where compact arrays were found to perform best at a fixed number of atoms~\cite{VanDamme2021}. For XXZ couplings, Begg \textit{et al.} found that squeezing in two-dimensional lattices with vacancies keeps growing with the number of atoms only below a threshold density of vacancies~\cite{Begg2026}, and Solfanelli \textit{et al.} related the squeezing on random interaction graphs to the spectral dimension of the graph~\cite{Solfanelli2026}. For Ising couplings, how much holes reduce the squeezing in two dimensions, which property of the array sets this reduction, and whether filling the holes is worth the atoms it loses, have remained open. In addition, the atoms lose coherence through dephasing and spontaneous emission, which limit the squeezing further.

In this work, we use the exact solution of the Ising dynamics to study how the arrangement of the atoms sets the optimal squeezing. We find that holes reduce the squeezing by more than the loss of atoms alone, while the number of atoms matters little. We trace both observations to two statistics of the couplings, the effective number of partners $\kappa$ and the spread of the local fields $\epsilon$, and we derive a formula $F(\kappa,\epsilon)$ that predicts the optimal squeezing and the optimal time from these two numbers, without fitted parameters. The formula is accurate for two-dimensional arrays with couplings that decay with distance, including arrays with holes and arrays of different shapes. From the formula, we propose design rules for the shape of an array and for filling its holes. We then extend the formula to dephasing and spontaneous emission, and we apply it to Rydberg-dressed couplings, where we compare it with the arrays of a recent experiment~\cite{Eckner2023}.

The paper is organized as follows. Section~\ref{sec:model} introduces the model and its exact solution. Section~\ref{sec:holes} quantifies how holes reduce the squeezing, and Sec.~\ref{sec:formula} derives the formula that explains it. Section~\ref{sec:design} turns the formula into design rules. Section~\ref{sec:decoherence} extends the formula to dephasing and spontaneous emission, Sec.~\ref{sec:softcore} applies it to Rydberg-dressed couplings and to the arrays of a recent experiment, and Sec.~\ref{sec:discussion} discusses the results.

% =====================================================================================
\section{Model and exact solution}
\label{sec:model}
In this section, we introduce the spin model, solve its dynamics exactly, and define the squeezing parameter used throughout this work. We consider atoms trapped in a two-dimensional array that interact through Ising couplings decaying with distance. Such couplings arise from Rydberg dressing~\cite{Zeiher2016,Borish2020,Hines2023,Eckner2023}, where they decay as the van der Waals interaction, and in crystals of trapped ions, where the decay can be tuned up to that of the dipolar interaction~\cite{Britton2012,Bohnet2016}. Because the Ising interaction commutes with every $\sigma^z_i$, the dynamics from a product state can be solved exactly for any arrangement of the atoms, including arrays with holes and position disorder.

The $N$ spin-$1/2$ atoms sit at positions $\mathbf r_i$ and evolve under the Ising Hamiltonian
\begin{equation}
  H=\sum_{i<j}J_{ij}\,\sigma^z_i\sigma^z_j ,
\label{eq:H}
\end{equation}
with couplings that decay as a power of the distance,
\begin{equation}
  J_{ij}=\frac{J_0}{|\mathbf r_i-\mathbf r_j|^{\alpha}},
\end{equation}
where we consider $\alpha=3$, which corresponds to dipolar interactions, and $\alpha=6$, which corresponds to van der Waals interactions. All spins start along $+x$, in the product state $|\psi_0\rangle=|{+}x\rangle^{\otimes N}$, and the interaction twists this product state into a squeezed state. We measure lengths in units of the lattice spacing and set $J_0=1$, so that the nearest-neighbor coupling is $J_{nn}=1$ and times are measured in units of $1/J_{nn}$.

Since every $\sigma^z_i$ commutes with $H$, each spin precesses about the $z$ axis in the field of its neighbors, and every correlator factorizes into a product of cosines~\cite{FossFeig2013,vandenWorm2013}. For a single spin,
\begin{equation}
  \langle\sigma^+_k\rangle=\tfrac12\prod_{j\neq k}\cos\theta_{kj},
\label{eq:exact}
\end{equation}
where the bond angle of atoms $k$ and $j$ is
\begin{equation}
  \theta_{kj}=2J_{kj}t .
\end{equation}
For two spins,
\begin{equation}
\begin{aligned}
  \langle\sigma^+_k\sigma^{\pm}_l\rangle&=\tfrac14\prod_{j\neq k,l}\cos(\theta_{kj}\pm\theta_{lj}),\\
  \langle\sigma^+_k\sigma^z_l\rangle&=\tfrac{i}{2}\sin\theta_{kl}\prod_{j\neq k,l}\cos\theta_{kj}.
\end{aligned}
\label{eq:exact2}
\end{equation}
Each cosine comes from one neighbor, whose up or down state shifts the precession phase of spin $k$ by $\pm\theta_{kj}$. The single-spin correlator determines the mean spin, and the two-spin correlators determine its fluctuations.

We combine the mean spin and its fluctuations into a single measure of squeezing, the Wineland parameter~\cite{Wineland1994},
\begin{equation}
  \xi^2=\frac{N\,\min_{\mathbf n\perp\langle\mathbf S\rangle}\mathrm{Var}(\mathbf n\cdot\mathbf S)}{|\langle\mathbf S\rangle|^2},
  \label{eq:xi}
\end{equation}
where $\mathbf S=\frac12\sum_i\bm\sigma_i$ is the collective spin and $\mathbf n$ is a unit vector perpendicular to the mean spin $\langle\mathbf S\rangle$, so that $\mathrm{Var}(\mathbf n\cdot\mathbf S)$ is the variance of the spin component along $\mathbf n$ and the minimum selects the squeezed direction. We also express the squeezing in decibels,
\begin{equation}
  \xi^2_{\rm dB}=10\log_{10}\xi^2 .
\end{equation}
Values $\xi^2<1$, or negative values in decibels, indicate squeezing, and throughout this work we study the squeezing at the optimal time.

By symmetry, the mean spin stays along $x$, and the squeezing takes place in the $y$--$z$ plane. The collective moments are sums over these correlators, and the minimal variance in Eq.~\eqref{eq:xi} follows from the $2\times2$ covariance matrix of $\Sy$ and $\Sz$, so that the exact squeezing can be computed for arrays of hundreds of atoms.

% =====================================================================================
\section{Squeezing in arrays with holes}
\label{sec:holes}
\begin{figure}[t]
  \includegraphics[width=\columnwidth]{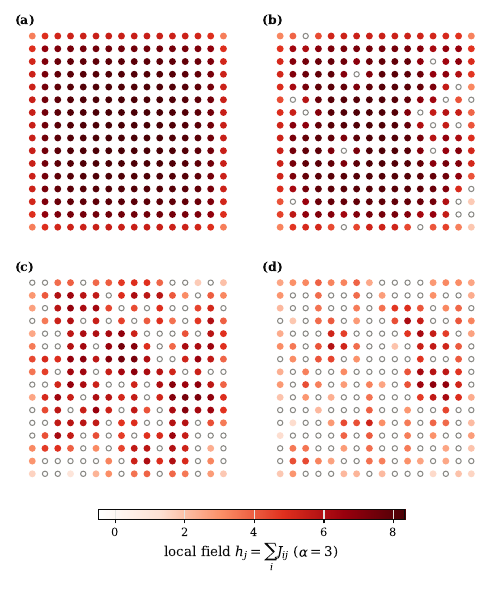}
  \caption{Square $16\times16$ arrays at (a) full filling, $N=256$, and at filling (b) $90\%$, $N=236$, (c) $70\%$, $N=177$, and (d) $50\%$, $N=133$, one random hole pattern each. Filled circles are atoms, colored by their local field $h_j=\sum_i J_{ij}$ for $\alpha=3$, and open circles are holes. Atoms at the edges and next to holes have weaker local fields.}
  \label{fig:arrays}
\end{figure}

Because random loading leaves holes that rearrangement does not always remove, we ask in this section how much the holes reduce the squeezing. Since the exact solution holds for any arrangement of atoms, we can compute the squeezing of such arrays directly. To see how incomplete filling affects the squeezing, we start from a $16\times16$ square grid and keep each atom with a probability equal to the filling, from the full grid down to half filling (Fig.~\ref{fig:arrays}). For each filling, we draw eight random hole patterns. Removing atoms weakens the local field $h_j=\sum_i J_{ij}$ of the atoms that remain, most strongly next to the holes, so the array becomes increasingly inhomogeneous as the filling drops.

\begin{figure}[t]
  \includegraphics[width=0.88\columnwidth]{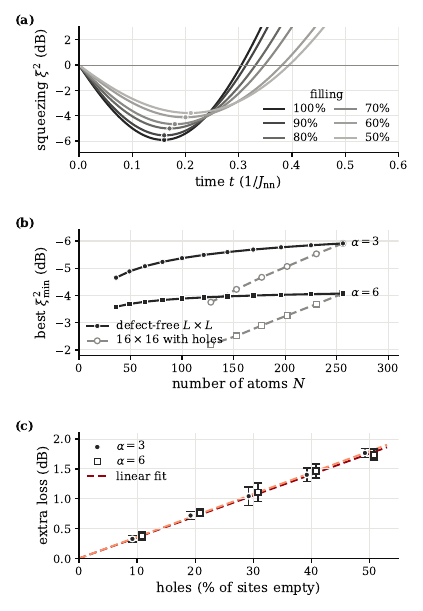}
  \caption{(a) Squeezing $\xi^2$ versus time on the $16\times16$ grid for $\alpha=3$, as the filling drops from full to half. Dots mark the optimum. (b) Best squeezing versus the number of atoms for defect-free $L\times L$ arrays (solid lines, filled markers) and for the $16\times16$ grid with holes (dashed lines, open markers, averaged over eight hole patterns), for $\alpha=3$ (circles) and $\alpha=6$ (squares). Stronger squeezing points up. (c) Extra loss of the arrays with holes, relative to a defect-free array with the same number of atoms, versus the fraction of empty sites. Error bars show the spread over the hole patterns, and dashed lines are linear fits.}
  \label{fig:holes}
\end{figure}

We measure the effect of holes by how much the best squeezing $\xi^2_{\min}$ degrades. Figure~\ref{fig:holes}(a) shows $\xi^2(t)$ for $\alpha=3$ as the filling drops from the full grid to half filling. Every curve has the same shape. The interaction first twists the spins and lowers the noise below the coherent-state level, and later the mean spin shrinks as the spins dephase, so $\xi^2$ passes through a minimum and rises again. As the filling drops, this minimum becomes shallower and moves to later times. The later optimum reflects the loss of partners, since each remaining atom collects random phases from fewer partners and therefore dephases more slowly. The shallower minimum has two possible origins, the smaller number of atoms and the disorder that the holes create.

Removing atoms reduces $N$, and smaller arrays squeeze less even without holes, so comparing an array with holes with the full grid mixes the two origins. To isolate the effect of the disorder, we compare each array with holes to a defect-free square array that holds the same number of atoms, whose squeezing we interpolate between defect-free $L\times L$ arrays of side length $L$ [Fig.~\ref{fig:holes}(b)]. The defect-free arrays give the squeezing that $N$ atoms reach when they are packed compactly. Their squeezing improves only slowly as the array grows, and even more slowly for $\alpha=6$, so the number of atoms matters little. With couplings that decay with distance, each atom twists mainly with its nearby partners, so atoms added far away change little, and the faster decay of the van der Waals interaction confines these partners to the nearest sites, so that the squeezing approaches its large-array value at even smaller sizes. The arrays with holes squeeze less than this reference at every filling and for both $\alpha$, so the holes reduce the squeezing by more than the loss of atoms alone.

For each filling, we measure how much less an array with holes squeezes than the defect-free array with the same number of atoms. We call this shortfall the extra loss and plot it against the fraction of empty sites in Fig.~\ref{fig:holes}(c). It grows roughly linearly with the fraction of holes, hardly depends on the particular hole pattern, and is nearly the same for dipolar and van der Waals couplings. This extra loss arises because every remaining atom loses some of its partners, and atoms next to holes lose more than the others, so the atoms have fewer partners than in a compact array and their local fields become unequal. Atoms with different local fields twist by different amounts, so no single direction is squeezed for all of them at once, and the collective noise rises.

Together, these results show that neither the number of atoms nor the filling alone sets the squeezing. What matters is how many partners each atom twists with and how equal the local fields are across the array.


% =====================================================================================
\section{A two-number formula for the optimum}
\label{sec:formula}
Section~\ref{sec:holes} has shown that holes reduce the squeezing by more than the loss of atoms alone, and that the number of atoms matters little. Both observations point to two properties of the array, how many partners each atom twists with and how equal the local fields are, but they do not tell us how these properties set the squeezing. In this section, we resolve this question and show that two statistics of the couplings determine the optimal squeezing of an array.

For each atom $j$, we define the local field $h_j=\sum_iJ_{ij}$, the sum of its couplings, and $q_j=\sum_iJ_{ij}^2$, the sum of its squared couplings, and we denote averages over the atoms by a bar, so that $\hb=N^{-1}\sum_jh_j$ is the mean local field and $\qb=N^{-1}\sum_jq_j$ the mean sum of squared couplings. The local field sets how strongly an atom twists, while $q_j$ sets how fast it dephases, because each partner adds a random phase whose variance grows with the squared coupling. The first number we need is
\begin{equation}
  \kappa=\frac{\hb^{\,2}}{\qb}.
\label{eq:kappa}
\end{equation}
It counts the effective partners each atom twists with. An atom that couples equally to $n$ partners has $h_j=nJ$ and $q_j=nJ^2$, so that $\kappa=n$, and couplings that decay with distance give $\kappa$ of the order of the number of nearby neighbors. The second number is the relative spread of the local fields,
\begin{equation}
  \epsilon=\frac{\overline{(h-\hb)^2}}{\hb^{\,2}}.
\label{eq:eps}
\end{equation}
In an infinite uniform array, every atom sees the same surroundings, all local fields are equal, and $\epsilon$ vanishes. Atoms at an edge miss the partners on one side, and atoms next to a hole miss the partners at the empty sites, so both edges and holes push some local fields below the mean and make $\epsilon$ grow.

To see how these two numbers set the squeezing, we measure the mean spin and the noise in units of their coherent-state values and write $\langle\Sx\rangle=NC/2$, $\mathrm{Var}\,\Sy=(N/4)(1+x)$, $\mathrm{Cov}(\Sy,\Sz)=(N/4)\,b$ and $\mathrm{Var}\,\Sz=N/4$. Here $C$ is the contrast, $b$ the correlation created by the twisting, and $x$ the noise added along $y$. Minimizing the variance in Eq.~\eqref{eq:xi} over the $y$--$z$ plane then gives the exact expression
\begin{equation}
  \xi^2=\frac{1+x/2-\sqrt{x^2/4+b^2}}{C^2}.
\label{eq:xiCbx}
\end{equation}
We now evaluate $C$, $b$ and $x$ when each atom has many weak bonds. Expanding $\ln\cos\theta\simeq-\theta^2/2$ turns every product of cosines in Eqs.~\eqref{eq:exact} and~\eqref{eq:exact2} into a Gaussian, $\prod_{j\neq k}\cos\theta_{kj}\simeq e^{-2t^2q_k}$, where $q_k$ is the sum of squared couplings of atom $k$. Replacing $q_k$ by its average gives the contrast,
\begin{equation}
  C^2=e^{-u},
\label{eq:C}
\end{equation}
where the phase variance that each atom accumulates from its partners is
\begin{equation}
  u=4t^2\qb .
\end{equation}
In the covariance, $\sin\theta_{kl}\simeq\theta_{kl}=2J_{kl}t$, and the sum over the partners of each atom is its local field, so that $b=2t\,\hb\,e^{-u/2}$. Multiplying and dividing by $\qb$ gives
\begin{equation}
  b^2=\kappa X,
\label{eq:b}
\end{equation}
where the factor
\begin{equation}
  X=u\,e^{-u}
\end{equation}
grows with the phase variance at early times and is suppressed by the loss of contrast at late times. In the variance, the two products of Eq.~\eqref{eq:exact2} become $e^{-u}e^{\mp M_{kl}}$, where $M_{kl}=4t^2\sum_jJ_{kj}J_{lj}$ measures the partners that atoms $k$ and $l$ share. Their difference is proportional to $\sinh M_{kl}$, and keeping only its first order, $\sinh M_{kl}\simeq M_{kl}$, which holds when two atoms share only a small part of their partners, gives $x\simeq(e^{-u}/N)\sum_{k\neq l}M_{kl}$. Summing over the shared partner first gives $\sum_{k\neq l}M_{kl}=4t^2\sum_j(h_j^2-q_j)$, and with $\overline{h^2}=\hb^{\,2}(1+\epsilon)$,
\begin{equation}
  x=X\big[\kappa(1+\epsilon)-1\big].
\label{eq:x}
\end{equation}
The twisting in Eq.~\eqref{eq:b} depends only on the mean local field $\hb$, whereas the noise in Eq.~\eqref{eq:x} grows with the mean square of the local field, $\hb^{\,2}(1+\epsilon)$. The spread $\epsilon$ therefore adds noise without adding twisting, which is why inhomogeneous arrays squeeze less, and $\kappa$ measures how strongly the twisting outweighs the dephasing.
\begin{figure}[b]
  \includegraphics[width=0.88\columnwidth]{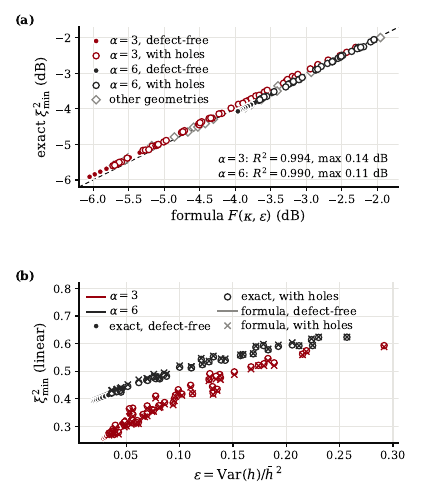}
  \caption{(a) Exact optimal squeezing versus the prediction of $F(\kappa,\epsilon)$ for defect-free square arrays of increasing size (filled) and $16\times16$ arrays with holes (open), for $\alpha=3$ (red) and $\alpha=6$ (black), and for the other geometries of Table~\ref{tab:accuracy} with $\alpha=3$, $4$ and $6$ (grey diamonds). The dashed line marks perfect agreement. (b) Optimal squeezing versus the inhomogeneity $\epsilon$, exact (circles) and from the formula (lines for defect-free arrays, crosses for arrays with holes), with colors as in (a). For $\alpha=3$, the arrays with holes extend down to $30\%$ filling, so that the arrays for both couplings cover the same range of $\epsilon$.}
  \label{fig:formula}
\end{figure}
\begin{figure}[t]
  \includegraphics[width=0.88\columnwidth]{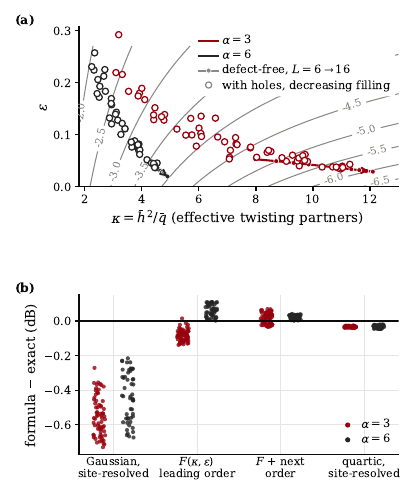}
  \caption{(a) Contours of $F$, in dB, in the plane of $\kappa$ and $\epsilon$, with defect-free arrays growing from $6\times6$ to $16\times16$ (lines, arrows in the direction of growth) and $16\times16$ arrays with holes at decreasing filling, down to $0.3$ for $\alpha=3$ and $0.5$ for $\alpha=6$ (open circles), for $\alpha=3$ (red) and $\alpha=6$ (black). (b) Difference between the predicted and the exact optimal squeezing, in dB, for the defect-free arrays and the arrays with holes of Table~\ref{tab:accuracy} at four levels of approximation, from a Gaussian expansion kept atom by atom, through $F$ and $F$ with its next-order correction, to a quartic expansion kept atom by atom.}
  \label{fig:plane}
\end{figure}
Inserting Eqs.~\eqref{eq:C}--\eqref{eq:x} into Eq.~\eqref{eq:xiCbx} and minimizing over time gives the optimal squeezing as a function of $\kappa$ and $\epsilon$ alone, which tells us how much squeezing an array can reach,
\begin{equation}
  \xi^2_{\min}\simeq F(\kappa,\epsilon)=\min_{u>0}\,e^{u}\Big[1+\frac{x}{2}-\sqrt{\frac{x^2}{4}+\kappa X}\Big].
\label{eq:F}
\end{equation}
The expression minimized in Eq.~\eqref{eq:F} weighs the twisting $\kappa X$ against the noise $x$ added along $y$ and the loss of contrast $e^{u}=1/C^2$, and it depends on time only through the phase variance $u$.

Foss-Feig \textit{et al.}~\cite{FossFeig2016} expanded the same exact solution in the system size. For homogeneous lattices, they found how the optimal squeezing scales with the system size and with $\alpha$, and they showed that, with open boundaries, the spread of the local fields makes the best squeezing of large systems independent of their size. This spread corresponds to our $\epsilon$. Our expansion is instead in the small angle of each single bond and keeps the finite number of effective partners $\kappa$, which yields a closed formula for finite arrays, including arrays with holes, edges and irregular shapes.

We now test the formula against the exact solution. Figure~\ref{fig:formula}(a) compares $F(\kappa,\epsilon)$ with the exact optimum for defect-free square arrays of increasing size and for $16\times16$ arrays with holes, for both $\alpha$. For $\alpha=3$, we extend the arrays with holes down to $30\%$ filling, so that their spread of local fields reaches that of the $\alpha=6$ arrays. The formula follows the exact result closely over the whole range, without any fitted parameter (Table~\ref{tab:accuracy}). It is equally accurate for other geometries, namely a smaller grid with holes, elongated rectangles with and without holes, arrays at low filling and arrays with position disorder, both for the couplings above and for $\alpha=4$.
Figure~\ref{fig:formula}(b) shows how the two numbers act separately. The squeezing degrades steadily as $\epsilon$ grows, and at the same $\epsilon$, the $\alpha=6$ arrays squeeze less than the $\alpha=3$ arrays because their atoms have fewer partners, that is, a smaller $\kappa$. The formula $F$ reproduces both the degradation with $\epsilon$ and the gap between the two couplings over the whole range of $\epsilon$.

Furthermore, we can also explain the behavior of the arrays with holes from this expression. Random holes reduce $\kappa$ roughly in proportion to the filling, since every atom loses partners, and they raise $\epsilon$ roughly in proportion to the fraction of holes, since atoms next to holes lose more partners than the others [Fig.~\ref{fig:plane}(a)]. Together, these two changes move the arrays with holes toward lower squeezing in the plane of $\kappa$ and $\epsilon$. Defect-free arrays move the other way as they grow, toward larger $\kappa$ and smaller $\epsilon$, because their atoms gain partners and the edges matter less. This also explains why the squeezing of the defect-free arrays in Fig.~\ref{fig:holes}(b) grows only slowly with size. As an array grows, $\kappa$ keeps increasing toward its value in an infinite array, which is finite because the couplings decay with distance. At the same time, $F$ depends ever more weakly on $\kappa$, as the widening spacing of its contours at large $\kappa$ in Fig.~\ref{fig:plane}(a) shows, so adding atoms improves the squeezing less and less.

The derivation of $F$ assumes that each single bond twists an atom by only a small angle and that two atoms share only a small part of their partners. In sparse arrays, these conditions hold only roughly, yet the formula remains accurate for every array in Table~\ref{tab:accuracy}, including arrays at low filling in which each atom has only a few effective partners. As the couplings decay more slowly, pairs of atoms share more of their partners, and the prediction moves from slightly less to slightly more squeezing than the exact result. It fails for all-to-all coupling, where pairs of atoms share all their partners and the formula overestimates the squeezing. For one-axis twisting, which is this limit, we find that the overestimate is small for a few atoms and grows steadily with the number of atoms. Rather than rely on the assumptions of the derivation, we establish the accuracy of $F$ for two-dimensional arrays with couplings that decay with distance by direct comparison with the exact solution.

The accuracy of $F$ is not accidental. Figure~\ref{fig:plane}(b) compares it with three other levels of the same expansion in the bond angle. The first keeps the Gaussian factor of every atom and every pair separately, including the terms from the bond between the two atoms of a pair, which the two-spin products exclude, but without the quartic term of $\ln\cos\theta$, which is of the same order. It gives a larger error than $F$, because it keeps some corrections of the next order but not others. The second adds to $F$ all corrections of the next order, namely the quartic terms, the single-bond terms and the variation of $q_k$ from atom to atom, written through further statistics of the couplings. The third keeps the Gaussian and quartic terms of every atom and every pair separately. Both complete next-order levels bring the prediction close to the exact result. $F$ is therefore the consistent leading term of this expansion, and its small remaining error comes from the next-order corrections that it neglects.

To prepare this squeezed state in practice, we also need to know how long the atoms should interact. Equation~\eqref{eq:F} contains this information, because its minimum is reached at a particular value $u^\ast$ of the phase variance. Since the phase variance grows with time as $u=4t^2\qb$, this value corresponds to the optimal interaction time,
\begin{equation}
  t^\ast=\frac12\sqrt{\frac{u^\ast}{\qb}}.
\label{eq:tstar}
\end{equation}
Because $u^\ast$ changes only slowly from one array to another, the optimal time is set mainly by $\qb$. Atoms with strong couplings to their partners therefore squeeze faster. Holes remove partners and lower $\qb$, which is why the optimum in Fig.~\ref{fig:holes}(a) moves to later times as the filling drops. The optimal time from Eq.~\eqref{eq:tstar} agrees closely with the exact one for every array (Table~\ref{tab:accuracy}).

\begin{table}
  \caption{Accuracy of $F(\kappa,\epsilon)$ against the exact optimal squeezing, with the error defined as $F$ minus the exact value. The mean, largest absolute and rms errors of the squeezing are in dB, and the column $t^\ast$ gives the rms relative error of the optimal time from Eq.~\eqref{eq:tstar} in percent. Negative values mean that the formula predicts slightly more squeezing than the exact solution. The $R^2$ is given only for whole families, because the defect-free arrays alone span too narrow a range for it to be meaningful. The other geometries are a defect-free $12\times12$ square, a $16\times16$ array at $40\%$ filling, a $14\times14$ array at $75\%$ filling, defect-free $6\times24$ and $4\times30$ rectangles, an $8\times20$ rectangle at $70\%$ filling, and a $12\times12$ square with position disorder, each for $\alpha=3$, $4$ and $6$.}
  \label{tab:accuracy}
  \begin{ruledtabular}
  \setlength{\tabcolsep}{4pt}
  \begin{tabular}{lcccccc}
   family & arrays & mean & max & rms & $t^\ast$ & $R^2$\\ \hline
   \multicolumn{7}{l}{$\alpha=3$}\\
   \quad defect-free & 11 & $-0.12$ & 0.14 & 0.12 & 4.8 & \\
   \quad with holes  & 56 & $-0.08$ & 0.13 & 0.08 & 3.8 & \\
   \quad all         & 67 & $-0.09$ & 0.14 & 0.09 & 4.0 & 0.994\\
   \multicolumn{7}{l}{$\alpha=6$}\\
   \quad defect-free & 11 & $+0.11$ & 0.11 & 0.11 & 1.7 & \\
   \quad with holes  & 40 & $+0.05$ & 0.10 & 0.05 & 1.0 & \\
   \quad all         & 51 & $+0.06$ & 0.11 & 0.07 & 1.2 & 0.990\\ \hline
   other geometries & 21 & $-0.01$ & 0.14 & 0.08 & 2.8 & 0.995\\
  \end{tabular}
  \end{ruledtabular}
\end{table}

% =====================================================================================
\section{Design rules}
\label{sec:design}
In the previous two sections, we showed that the squeezing of an array is set by two numbers, the effective number of partners $\kappa$ and the spread of the local fields $\epsilon$. When building an array, one controls $\kappa$ and $\epsilon$ through two choices, the shape in which a given number of atoms is arranged and whether the holes left by loading are filled. In this section, we use the formula $F(\kappa,\epsilon)$ to guide the choice of shape and the decision to fill the holes, and we propose two design rules.

\begin{figure}[t]
  \includegraphics[width=0.88\columnwidth]{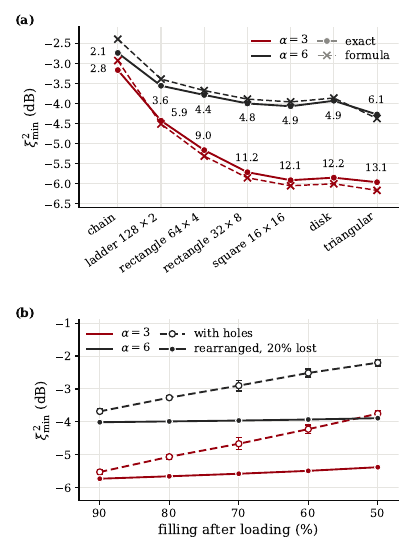}
  \caption{(a) Best squeezing for arrays of different shapes at a fixed number of atoms, $N\approx256$, exact (filled circles, solid lines) and from the formula $F(\kappa,\epsilon)$ (crosses, dashed lines), for $\alpha=3$ (red) and $\alpha=6$ (black). The shapes are ordered by their $\kappa$ for $\alpha=3$, and the number next to each exact point gives its $\kappa$. The disk holds $N=253$ atoms. (b) Best squeezing versus the filling of the $16\times16$ grid after loading, for the arrays with holes (open circles, dashed lines, averaged over eight hole patterns, with error bars showing the spread) and for compact arrays that hold a fifth fewer atoms than the arrays with holes, as left by a rearrangement that loses some atoms (filled circles, solid lines), for $\alpha=3$ (red) and $\alpha=6$ (black). The compact arrays are interpolated between defect-free $L\times L$ arrays, as in Fig.~\ref{fig:holes}(b).}
  \label{fig:design}
\end{figure}

We first compare arrays of different shapes that hold the same number of atoms at the same nearest-neighbor spacing [Fig.~\ref{fig:design}(a)]. For dipolar and van der Waals couplings alike, we find that the squeezing improves overall with $\kappa$, from the chain, through the ladder and the elongated rectangles, to the square, the disk and the triangular lattice. To understand the ranking of the shapes, we look at how many close partners each atom has in each shape. The triangular lattice gives every atom six nearest neighbors, against four in the square, and elongated shapes place a larger share of the atoms at an edge, where some of their partners are missing. We see that the number of nearest neighbors and the share of atoms at an edge together set $\kappa$.

When two shapes have similar values of $\kappa$, we find that the spread of the local fields decides which of them squeezes more. The disk, for instance, has a slightly larger $\kappa$ than the square but squeezes slightly less, because its curved edge makes the local fields of the atoms near the boundary more unequal than in the square.

We also find that the choice of shape changes the squeezing much more for dipolar than for van der Waals couplings. The slowly decaying dipolar coupling links each atom appreciably to many partners in every direction, so stretching the array into a ladder or a chain removes many of these partners. The rapidly decaying van der Waals coupling links each atom mainly to its nearest neighbors in any shape, so stretching the array removes fewer of its partners. As a result, the squeezing falls much further from the triangular lattice to the chain for $\alpha=3$ than for $\alpha=6$.

Based on this comparison of shapes, we propose a first design rule. For a fixed number of atoms and a fixed nearest-neighbor spacing, the array should be as compact and as highly coordinated as possible, because many close partners and few atoms at an edge give a large $\kappa$ and a small $\epsilon$, which the formula turns into more squeezing. The triangular lattice is therefore the best of the shapes we considered, although its advantage over the square is small, and the choice of shape matters most for dipolar couplings.

The formula $F$ reproduces the ranking of the shapes for dipolar and van der Waals couplings alike. The difference between the formula and the exact squeezing stays small for all two-dimensional shapes and grows only for the chain, for which the formula predicts less squeezing than the exact solution, although each atom of the chain has about as many effective partners as in the sparsest two-dimensional arrays of Table~\ref{tab:accuracy}. The number of effective partners alone therefore does not mark where the formula holds.

For the second choice, we ask whether the atoms of an array with holes should be rearranged into a compact array without holes~\cite{Barredo2016,Endres2016}. The extra loss in Fig.~\ref{fig:holes}(c) compares an array with holes with a compact array of the same number of atoms, so it is the squeezing that rearranging recovers when no atoms are lost. We therefore conclude that rearranging increases the squeezing at every filling below one, and that the gain grows roughly linearly with the fraction of holes for dipolar and van der Waals couplings alike. In the plane of $\kappa$ and $\epsilon$ in Fig.~\ref{fig:plane}(a), we see that rearranging moves an array with holes back toward larger $\kappa$ and smaller $\epsilon$, where the formula predicts more squeezing. In practice, rearranging atoms often loses some of them~\cite{Endres2016}. Because the squeezing of compact arrays depends only weakly on the number of atoms [Fig.~\ref{fig:holes}(b)], as expected for Ising couplings that decay faster than $1/r^D$ in $D$ dimensions~\cite{FossFeig2016}, this loss of atoms turns out to matter little. A compact array with a fifth fewer atoms still squeezes clearly more than the array with holes from which it was made, for dipolar and van der Waals couplings alike and at moderate as well as low filling [Fig.~\ref{fig:design}(b)]. From this comparison of arrays with holes and compact arrays, we propose a second design rule. After loading, the atoms should be rearranged into a compact array without holes, even when the rearrangement loses some of them, because removing the holes raises $\kappa$ and lowers $\epsilon$, whereas losing atoms barely changes the squeezing of a compact array. In one dimension, Van Damme \textit{et al.} found that compact arrays perform best at a fixed number of atoms~\cite{VanDamme2021}, and our results extend this conclusion to two dimensions and explain it through $\kappa$ and $\epsilon$.

Combining the two design rules with the formula, we arrive at a simple recipe for designing an array. From the atom positions and the coupling law, one computes the local fields $h_j$ and the sums of squared couplings $q_j$, then $\kappa$ and $\epsilon$, and reads off the best squeezing from Eq.~\eqref{eq:F} and the optimal interaction time from Eq.~\eqref{eq:tstar}. No simulation of the dynamics is needed, and the recipe applies to two-dimensional arrays with couplings that decay with distance, including arrays with position disorder.

% =====================================================================================
\section{Decoherence}
\label{sec:decoherence}
So far, we have treated the dynamics as purely coherent. In experiments, the atoms also lose coherence, through dephasing, which randomizes the phase of each spin without changing its populations, and through spontaneous emission, which in addition returns the spins from $|{\uparrow}\rangle$ to $|{\downarrow}\rangle$. In Rydberg-dressed arrays, for example, the admixture of the Rydberg state brings both the decay of that state~\cite{Henkel2010,Pupillo2010} and additional phase noise~\cite{Goldschmidt2016,Zeiher2016}. In this section, we ask how much of each process the squeezing tolerates, and we extend the formula $F(\kappa,\epsilon)$ so that it predicts the optimal squeezing in the presence of both. We describe dephasing at rate $\gamma$ and spontaneous emission at rate $\Gamma$ with the master equation
\begin{equation}
\begin{aligned}
  \dot\rho={}&-i[H,\rho]+\frac{\gamma}{2}\sum_k\big(\sigma^z_k\rho\,\sigma^z_k-\rho\big)\\
  &+\Gamma\sum_k\Big(\sigma^-_k\rho\,\sigma^+_k-\tfrac12\big\{\sigma^+_k\sigma^-_k,\rho\big\}\Big),
\end{aligned}
\label{eq:master}
\end{equation}
where the dephasing term makes the coherence of each spin decay as $e^{-\gamma t}$, and the emission term takes each spin from $|{\uparrow}\rangle$ to $|{\downarrow}\rangle$. For $\gamma=\Gamma=0$ the dynamics reduces to that of the Ising Hamiltonian alone.

\subsection{Dephasing}
\begin{figure}[t]
  \includegraphics[width=0.88\columnwidth]{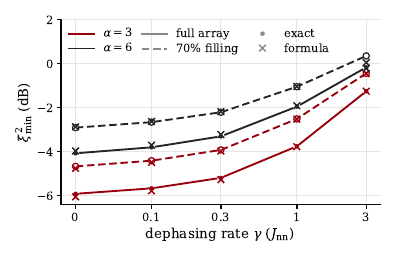}
  \caption{Best squeezing versus the dephasing rate $\gamma$ for the full $16\times16$ array (solid lines) and at $70\%$ filling (dashed lines), exact (circles) and from the formula (crosses), for $\alpha=3$ (red) and $\alpha=6$ (black). The leftmost points are without dephasing.}
  \label{fig:dephasing}
\end{figure}

We first consider dephasing. Because the dephasing term commutes with the Ising Hamiltonian, it leaves the populations unchanged and only multiplies every single-spin coherence by $e^{-\gamma t}$. We now follow these factors through the quantities $C$, $b$ and $x$ of Sec.~\ref{sec:formula}. The contrast $C=2\langle\Sx\rangle/N$ is a sum of single-spin coherences $\langle\sigma^+_k\rangle$, so that
\begin{equation}
  C_\gamma=e^{-\gamma t}\,C .
\end{equation}
The twisting correlation $b=\mathrm{Cov}(\Sy,\Sz)/(N/4)$ is a sum of the correlators $\langle\sigma^+_k\sigma^z_l\rangle$, each of which contains one coherence and one population, and the populations do not decay. Hence
\begin{equation}
  b_\gamma=e^{-\gamma t}\,b .
\end{equation}
In the variance, $\mathrm{Var}\,\Sy=N/4+\frac18\sum_{k\neq l}(P^-_{kl}-P^+_{kl})$, where $P^\pm_{kl}=\prod_{j\neq k,l}\cos(\theta_{kj}\pm\theta_{lj})$ are the products in Eq.~\eqref{eq:exact2}, the constant $N/4$ comes from $(\sigma^y_k)^2=1$ and does not decay, whereas the products $P^\pm_{kl}$ come from the two-spin correlators $\langle\sigma^+_k\sigma^\pm_l\rangle$, which contain two coherences. Only the excess variance decays, and the added noise $x=4(\mathrm{Var}\,\Sy-N/4)/N$ becomes
\begin{equation}
  x_\gamma=e^{-2\gamma t}\,x .
\end{equation}
The variance $\mathrm{Var}\,\Sz=N/4$ stays unchanged. Inserting $C_\gamma$, $b_\gamma$ and $x_\gamma$ into Eq.~\eqref{eq:xiCbx} gives the squeezing with dephasing,
\begin{equation}
  \xi^2=e^{u+2\gamma t}\Big[1+\frac{e^{-2\gamma t}x}{2}-\sqrt{\frac{e^{-4\gamma t}x^2}{4}+e^{-2\gamma t}\kappa X}\Big],
  \label{eq:xi_deph}
\end{equation}
with $u$, $X$ and $x$ as in Eqs.~\eqref{eq:C}--\eqref{eq:x}. Minimizing over time gives the formula with dephasing, without any new parameter. We find that the formula with dephasing follows the exact optimum closely over the whole range of dephasing rates in Fig.~\ref{fig:dephasing}, for dipolar and van der Waals couplings and for full as well as partially filled arrays.

We see in Fig.~\ref{fig:dephasing} that the squeezing degrades steadily as the dephasing rate grows. Arrays with holes lose about as much squeezing as full arrays at the same dephasing rate, but because they start with less squeezing, they lose all of it at lower dephasing rates. From these curves, we propose a practical rule. The dephasing should remain weak over the optimal time, $\gamma t^\ast\ll1$, where the squeezing remains close to its value without dephasing. For the arrays of Fig.~\ref{fig:dephasing}, whose optimal times are a fraction of $1/J_{nn}$, this requires a dephasing rate well below the nearest-neighbor coupling.

\subsection{Spontaneous emission}
We now turn to spontaneous emission, which takes each spin from $|{\uparrow}\rangle$ to $|{\downarrow}\rangle$ at rate $\Gamma$. Unlike dephasing, spontaneous emission changes the populations, so it does not simply multiply the coherent result by decay factors. The dynamics nevertheless remains exactly solvable. Because the Ising Hamiltonian is diagonal in the $\sigma^z$ basis, the populations of different atoms decay classically and independently of each other~\cite{FossFeig2013}. Each partner of an atom then imprints a phase that depends on when that partner decays, and averaging over these random decay times replaces each cosine in Eqs.~\eqref{eq:exact} and~\eqref{eq:exact2} by a function of the coupling, the time and $\Gamma$. We checked the resulting correlators against a direct numerical solution of the master equation for small arrays. The same exact treatment of decoherence in Ising dynamics was used by Young \textit{et al.} for Rydberg-dressed couplings~\cite{Young2023}.

Spontaneous emission also changes the direction of the mean spin. As the atoms decay toward $|{\downarrow}\rangle$, the mean spin tilts toward $-z$, and the decaying populations make it rotate in the $x$--$y$ plane. We therefore compute the squeezing from the full covariance matrix of the collective spin, in the plane perpendicular to the mean spin, as in Eq.~\eqref{eq:xi}.

\begin{figure}[t]
  \includegraphics[width=0.88\columnwidth]{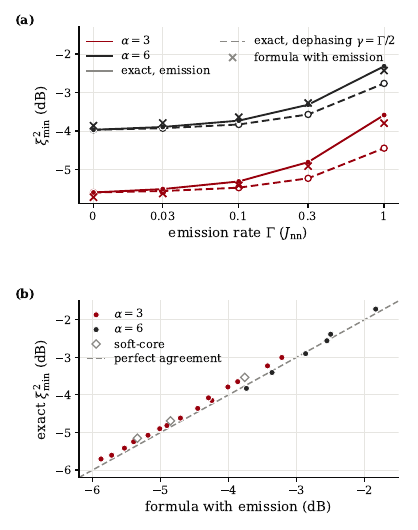}
  \caption{(a) Best squeezing versus the emission rate $\Gamma$ for a $12\times12$ array, exact with spontaneous emission (solid lines, filled circles), exact with dephasing at the same loss of coherence, $\gamma=\Gamma/2$ (dashed lines, open circles), and from the formula with emission (crosses), for $\alpha=3$ (red) and $\alpha=6$ (black). The leftmost points are without decoherence. (b) Formula with emission versus the exact optimal squeezing for square arrays of different sizes, arrays with holes, an elongated rectangle, the triangular lattice and Rydberg-dressed couplings, which saturate at short distances, each at three emission rates. The dashed line marks perfect agreement.}
  \label{fig:emission}
\end{figure}

Figure~\ref{fig:emission}(a) compares spontaneous emission with dephasing at the same loss of single-spin coherence, $\gamma=\Gamma/2$. We find that spontaneous emission reduces the squeezing more than dephasing at every emission rate, and the difference grows with $\Gamma$. The reason is that a decaying partner changes the phase it imprints on an atom at a random time, which adds phase noise beyond the loss of single-spin coherence. Describing spontaneous emission as dephasing with $\gamma=\Gamma/2$ therefore overestimates the squeezing, except at very weak emission.

\subsection{Formula with spontaneous emission}
The exact solution with spontaneous emission requires the full coupling matrix. To keep the simplicity of $F(\kappa,\epsilon)$, we extend the formula by giving each of its ingredients the decay statistics of a single spin. A spin that starts in a superposition of $|{\uparrow}\rangle$ and $|{\downarrow}\rangle$ and decays at rate $\Gamma$ has a random $z$ component $z(t)$, and the phase it imprints on a partner grows with the time integral $Z=\int_0^tz\,dt'$. Three averages of this random process enter the formula, the mean $m_1$ of $Z$, its variance $v$ and its covariance $w$ with the final value $z(t)$. Keeping the first two cumulants of the phase, as for the cosine products in Sec.~\ref{sec:formula}, a partner with coupling $J$ contributes the average
\begin{equation}
  \mathbb E\big[e^{2iJZ}\big]\simeq e^{2iJm_1-2J^2v}.
\end{equation}
Multiplying these averages over all partners of atom $k$ replaces $J$ by the local field $h_k$ in the first term and $J^2$ by $q_k$ in the second. Together with the damping of the coherence by the decay of atom $k$ itself, the coherence of atom $k$ becomes
\begin{equation}
  \langle\sigma^+_k\rangle\simeq\tfrac12\,e^{-\Gamma t/2}\,e^{-u_k/2}\,e^{2ih_km_1},
\end{equation}
where $u_k=4q_kv$ is the phase variance of atom $k$, which reduces to $4t^2q_k$ without emission. The new ingredient is the phase $2h_km_1$. Because decaying partners are biased toward $|{\downarrow}\rangle$, their mean phase no longer vanishes, and each atom precesses at a rate set by its own local field. The mean phase $2\hb m_1$ only rotates the mean spin in the $xy$ plane, whereas the spread of the local fields makes the atoms drift apart in phase. Averaging over the atoms, whose local fields have the variance $\epsilon\hb^{\,2}$, and keeping again the first two cumulants, we obtain
\begin{equation}
  \frac1N\sum_ke^{2i(h_k-\hb)m_1}\simeq D,
\end{equation}
where $D=\exp(-2m_1^2\hb^{\,2}\epsilon)$. With $u_k$ replaced by its mean $u=4\qb v$, the contrast becomes
\begin{equation}
  C_\Gamma=e^{-\Gamma t/2}\,D\,e^{-u/2}.
\end{equation}
For the twisting correlation, partner $l$ enters $\langle\sigma^+_k\sigma^z_l\rangle$ through both its phase and its own final $z$ component. To first order in the coupling,
\begin{equation}
  \mathbb E\big[z(t)\,e^{2iJZ}\big]\simeq e^{2iJm_1}\big(\bar z+2iJw\big),
\end{equation}
where $\bar z=e^{-\Gamma t}-1$ is the mean $z$ component of a decaying spin. The term with $\bar z$ is the product of the mean values of the two spins and drops out of the covariance, whereas the term with $w$ is the correlation. The bond angle $2J_{kl}t$ of Sec.~\ref{sec:formula} is thus replaced by $2J_{kl}w$, and summing over the partners gives
\begin{equation}
  b_\Gamma=2w\,\hb\;e^{-\Gamma t/2}\,D\,e^{-u/2}.
\end{equation}
The same steps applied to the two-spin products, which carry the coherence of two atoms, give the added noise
\begin{equation}
  x_\Gamma=e^{-\Gamma t}D^2X\big[\kappa(1+\epsilon)-1\big],
\end{equation}
where $X=u\,e^{-u}$. Without emission, $m_1=0$, $v=t^2$ and $w=t$, and these expressions reduce to Eqs.~\eqref{eq:C}--\eqref{eq:x}.

As in the exact solution, the mean spin tilts toward $-z$, because its $z$ component is $\bar z$. In units of $N/2$, the mean spin has the length $(C_\Gamma^2+\bar z^2)^{1/2}$ and makes an angle $\beta$ with the $xy$ plane, with
\begin{equation}
  \cos^2\!\beta=\frac{C_\Gamma^2}{C_\Gamma^2+\bar z^2}.
\end{equation}
We minimize the variance in the plane perpendicular to the mean spin, which is spanned by the in-plane direction $y'$ perpendicular to the mean spin and the tilted axis $n$. Along $n$, the fluctuations combine the population noise $1-\bar z^2$ of each spin with the in-plane noise $1-C_\Gamma^2$ along the mean spin, which we treat as uncorrelated, so that the normalized variance is
\begin{equation}
  V_n=\cos^2\!\beta\,\big(1-\bar z^2\big)+\sin^2\!\beta\,\big(1-C_\Gamma^2\big).
\end{equation}
Only the $z$ part of $n$ correlates with $y'$, so the twisting correlation along $n$ is $b_\Gamma\cos\beta$. The minimal normalized variance is then
\begin{equation}
  V_{\min}=\frac{1+x_\Gamma+V_n}{2}-\sqrt{\Big(\frac{1+x_\Gamma-V_n}{2}\Big)^2+b_\Gamma^2\cos^2\!\beta},
\end{equation}
and the squeezing parameter is
\begin{equation}
  \xi^2=\frac{V_{\min}}{C_\Gamma^2+\bar z^2}.
\end{equation}
Minimizing over time gives the formula with emission, which contains no fitted parameter and depends on the array only through $\kappa$, $\epsilon$ and $\qb$, where $\qb$ sets the time scale on which the emission rate $\Gamma$ acts.

Figure~\ref{fig:emission}(b) tests the formula with emission against the exact solution for square arrays of different sizes, arrays with holes, an elongated rectangle, the triangular lattice and Rydberg-dressed couplings, which saturate at short distances, each at weak, moderate and strong emission. We find that the formula follows the exact optimum closely in every case and becomes slightly optimistic only at the strongest emission. In Fig.~\ref{fig:emission}(a), we see that the formula with emission also captures the difference between spontaneous emission and dephasing on the $12\times12$ array, which the description by dephasing with $\gamma=\Gamma/2$ misses.

% =====================================================================================
\section{Rydberg-dressed (soft-core) couplings}
\label{sec:softcore}
So far, we have studied couplings that decay as a power law of the distance. Rydberg dressing, proposed in Ref.~\cite{Gil2014} and used in recent squeezing experiments with neutral atoms~\cite{Hines2023,Eckner2023}, produces couplings of a different shape. Dressing admixes a small fraction of a Rydberg state into the ground state of each atom, and the Rydberg blockade~\cite{Jaksch2000,Lukin2001} forbids two nearby atoms from being excited together. The coupling therefore saturates at short distances and takes the soft-core form~\cite{JohnsonRolston2010,Henkel2010,Pupillo2010,Honer2010}
\begin{equation}
  J(r)=\frac{V_0}{1+(r/R_c)^6},
  \label{eq:softcore}
\end{equation}
where $V_0$ is the coupling at short distances and $R_c$ the soft-core radius. Within the soft-core radius the coupling is nearly constant, and beyond it the coupling decays as $1/r^6$, like the van der Waals interaction between Rydberg atoms~\cite{Saffman2010}. Young \textit{et al.} showed that soft-core couplings squeeze more as more atoms fall within the soft-core radius~\cite{Young2023}. In this section, we use the formula $F(\kappa,\epsilon)$ to explain this growth through $\kappa$ and $\epsilon$, and we test the formula on the arrays of a recent experiment.

\begin{figure}[t]
  \includegraphics[width=0.88\columnwidth]{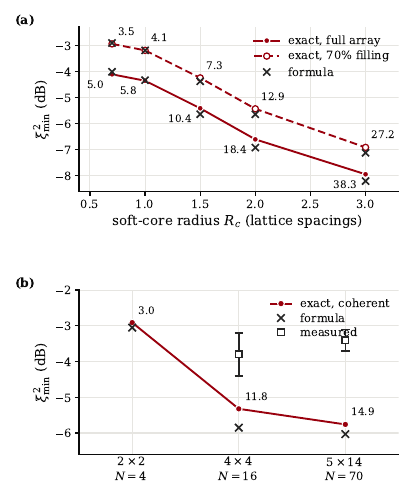}
  \caption{Squeezing with the Rydberg-dressed (soft-core) couplings of Eq.~\eqref{eq:softcore}. (a) Optimal squeezing $\xi^2_{\min}$ of a $16\times16$ square array as a function of the soft-core radius $R_c$, in units of the lattice spacing, for the full array (solid line, filled circles) and for one random pattern at $70\%$ filling (dashed line, open circles). Circles show the exact results and crosses the formula $F(\kappa,\epsilon)$. (b) Optimal squeezing for the fully filled array layouts of Eckner \textit{et al.}~\cite{Eckner2023}, with their fitted soft-core radius of $4.9$ optical-lattice sites. The $2\times2$ and $5\times14$ arrays have spacings of three and two lattice sites along the two axes, and the $4\times4$ array has a spacing of two sites. The line with circles shows the exact coherent squeezing, the crosses show the formula, and the squares with error bars show the squeezing measured in Ref.~\cite{Eckner2023}. In both panels, the number next to each exact point gives its $\kappa$.}
  \label{fig:softcore}
\end{figure}

We first follow the squeezing of a full $16\times16$ square array and of the same array at $70\%$ filling as the soft-core radius grows [Fig.~\ref{fig:softcore}(a)]. When the soft-core radius is smaller than the lattice spacing, every partner of an atom lies on the $1/r^6$ tail, and the array squeezes like the power law with $\alpha=6$. As the soft-core radius exceeds the lattice spacing, more and more partners fall inside the soft core, where they all share nearly the same coupling. We see that the effective number of partners $\kappa$ then grows roughly with the number of atoms inside the soft core, while the spread of the local fields $\epsilon$ stays small, and the squeezing improves strongly. In the language of the formula, the growth found by Young \textit{et al.} is a growth of $\kappa$.

Holes reduce the squeezing at every soft-core radius, as they do for power-law couplings, and we see that they remove a similar amount of squeezing at every radius. At every radius, the holes remove the same fraction of the partners inside the soft core, so they lower $\kappa$ by about the filling fraction. Their effect on $\epsilon$, by contrast, weakens as the soft core grows. When the soft core holds many atoms, each atom sums its local field over many partners, the random positions of the holes average out, and the local fields remain nearly equal.

The formula follows the exact results closely when the soft-core radius is small, and it predicts slightly more squeezing than the exact result when the radius is large. A large soft core makes the couplings between an atom and its nearby partners nearly equal, so pairs of atoms share most of their partners. The array then approaches all-to-all coupling, where the formula overestimates the squeezing, as we noted in Sec.~\ref{sec:formula}.

We now test the formula on an experiment. Eckner \textit{et al.} generated squeezing with Rydberg dressing in two-dimensional arrays of atoms and described the measured interactions with the soft-core form of Eq.~\eqref{eq:softcore}~\cite{Eckner2023}. We take their fitted soft-core radius and their array layouts and compute the exact coherent squeezing and the formula [Fig.~\ref{fig:softcore}(b)]. In these layouts, the soft-core radius is about two spacings of the array, so the arrays sit in the regime where the squeezing grows quickly with the soft-core radius. The formula reproduces the exact coherent squeezing closely. It is least accurate for the $4\times4$ array, where the soft core covers almost the whole array, so that pairs of atoms share most of their partners. The $2\times2$ array is also close to all-to-all coupling, but each atom has only three partners, and in this regime the overestimate of the formula stays small, as for one-axis twisting with a few atoms. The time $t^\ast$ of Eq.~\eqref{eq:tstar} also lands close to the exact optimal dressing time, so the formula gives both the best squeezing and the dressing time at which it occurs.

The measured squeezing lies well below this coherent prediction. Eckner \textit{et al.} found a similar gap against their own model and attributed it mainly to unitary dynamics beyond the two-body Ising description, arising from their strong dressing, with collective dissipation as a further possibility in the largest arrays~\cite{Eckner2023}. The contrast helps to test the role of decoherence. For the $4\times4$ array, the contrast that Eckner \textit{et al.} measured for the squeezed state near the optimal time is close to the contrast that the coherent Ising dynamics predicts at the optimum, although the measurement includes the atoms lost during dressing. Dephasing and spontaneous emission lower the contrast below its coherent value, as we showed in Sec.~\ref{sec:decoherence}, and atom loss lowers it as well. This agreement therefore suggests that these processes play at most a minor role near the optimal time, and that the missing squeezing appears mainly as excess variance of the squeezed spin component, consistent with dynamics beyond the Ising model rather than decoherence as the main limit.

Finally, the squeezing depends on the soft-core radius only through its ratio to the spacing of the array, because the coupling of Eq.~\eqref{eq:softcore} depends on distance only through $r/R_c$, and $V_0$ only sets the time scale. When building an array, one can therefore raise $\kappa$ either with a larger soft-core radius, through the choice of Rydberg state and the detuning of the dressing laser, or with a smaller spacing at fixed dressing. Van Damme \textit{et al.} studied random filling in one-dimensional arrays with a soft-core radius large compared with the spacing~\cite{VanDamme2021}. In this regime, each atom has many partners inside the soft core, so the holes lower $\kappa$ but leave the local fields nearly equal, as we found for the square array.

\section{Discussion}
\label{sec:discussion}
In this work, we studied how the arrangement of atoms in a two-dimensional array sets the spin squeezing generated by Ising couplings. Using the exact solution of the dynamics, we found that holes reduce the squeezing by more than the loss of atoms alone, and that the number of atoms matters little. We traced both observations to two statistics of the couplings, the effective number of partners $\kappa$ and the spread of the local fields $\epsilon$, and we derived a formula $F(\kappa,\epsilon)$ that predicts the optimal squeezing and the optimal time from these two numbers alone. A larger $\kappa$ gives more squeezing and a larger $\epsilon$ gives less, so an array squeezes best when each atom couples to many partners and all atoms feel nearly the same local field.

The formula turns the design of an array into a short calculation. From the coupling matrix of a proposed array, one computes $\kappa$ and $\epsilon$, reads off the optimal squeezing and the time at which it occurs, and compares candidate arrays without simulating the dynamics. From this calculation, we proposed two design rules. A given number of atoms should be arranged in an array that is as compact and as highly coordinated as possible, and the holes left by loading should be filled by rearranging the atoms, even when the rearrangement loses some of them. The formula extends to dephasing and spontaneous emission without any fitted parameter. With it, we found that the dephasing should remain weak over the optimal time, $\gamma t^\ast\ll1$, and that spontaneous emission removes more squeezing than dephasing at the same loss of single-spin coherence, because a decaying partner adds phase noise. For Rydberg-dressed couplings, the formula explains why a larger soft-core radius, or a smaller spacing of the array, improves the squeezing, and it reproduces the coherent prediction for the arrays of a recent experiment, whose measured squeezing lies well below that prediction.

The formula is accurate for two-dimensional arrays in which two atoms share only a small fraction of their partners, as for the power-law couplings studied here. For soft-core couplings with a large soft-core radius, and in the limit of all-to-all coupling, pairs of atoms share most of their partners, and the formula overestimates the squeezing. In that regime, the exact solution remains available, since it holds for any coupling matrix. Our analysis is also restricted to Ising couplings, for which the dynamics from a product state can be solved exactly for any arrangement of the atoms.

Several extensions remain open. Dipolar exchange couplings of XY type, as realized with Rydberg atoms in Ref.~\cite{Bornet2023}, do not conserve the individual $\sigma^z_i$, so the dynamics can no longer be solved exactly, and approximate methods such as the discrete truncated Wigner approximation~\cite{Schachenmayer2015,Truong2026}, which was applied to squeezing with spin-exchange couplings in Ref.~\cite{Perlin2020}, are needed to test whether $\kappa$ and $\epsilon$ also govern the squeezing there. The formula could also guide a search for optimized array geometries beyond the regular lattices and simple shapes considered here. Finally, extending the model to the strong dressing used in experiments, where the dynamics goes beyond two-body Ising couplings, and including decoherence and measurement imperfections would show how much of the gap between the coherent prediction and the measured squeezing each of these effects explains.

\begin{acknowledgments}
We thank Identity Quantum Computing for providing the computing hardware and for funding this work.
Parts of the simulation and analysis code, and the editing of the manuscript, were carried out with the assistance of an AI model (Claude, Anthropic). The authors designed the study, checked all code, results and text, and take full responsibility for the content of this article.
The simulation code, analysis scripts and data are available on GitHub~\cite{CodeRepo} and Zenodo~\cite{CodeZenodo}.
\end{acknowledgments}

\appendix
\section{Exact solution and its validation}
\label{app:validation}
All exact results in this work are evaluated from closed-form correlators, so their reliability rests on these expressions and on their numerical implementation. In this appendix, we collect the expressions that we evaluate and describe how we validated them. The collective moments are sums of the correlators of Eqs.~\eqref{eq:exact} and~\eqref{eq:exact2} over atoms and pairs of atoms. Because $\langle\sigma^+_k\rangle$ is real, the mean spin points along $x$, and $\langle\Sy\rangle=\langle\Sz\rangle=0$, so that its length is
\begin{equation}
  \langle\Sx\rangle = \tfrac12\sum_k\prod_{j\neq k}\cos\theta_{kj}.
\end{equation}
Writing $\sigma^y_k\sigma^y_l$ in terms of ladder operators expresses its expectation value through the two products of the first correlator in Eq.~\eqref{eq:exact2}, whereas $\langle\sigma^z_k\sigma^z_l\rangle$ vanishes at all times because the populations do not evolve, so that $V_{zz}\equiv\mathrm{Var}\,\Sz=N/4$ keeps its initial value. The variance along $y$ is
\begin{equation}
\begin{split}
  V_{yy}\equiv\mathrm{Var}\,\Sy &= \tfrac N4+\tfrac18\sum_{k\neq l}\Big[\prod_{j\neq k,l}\cos(\theta_{kj}-\theta_{lj})\\
  &\qquad-\prod_{j\neq k,l}\cos(\theta_{kj}+\theta_{lj})\Big],
\end{split}
\end{equation}
and the covariance follows from the imaginary part of the second correlator in Eq.~\eqref{eq:exact2},
\begin{equation}
  V_{yz}\equiv\mathrm{Cov}(\Sy,\Sz) = \tfrac14\sum_{k\neq l}\sin\theta_{kl}\prod_{j\neq k,l}\cos\theta_{kj}.
\end{equation}
The smaller eigenvalue of the covariance matrix of $\Sy$ and $\Sz$ is
\begin{equation}
  V_{\min} = \frac{V_{yy}+V_{zz}}{2}-\sqrt{\Big(\frac{V_{yy}-V_{zz}}{2}\Big)^2+V_{yz}^2},
\end{equation}
which is the minimum in Eq.~\eqref{eq:xi}, so that $\xi^2=NV_{\min}/\langle\Sx\rangle^2$. These expressions hold for any coupling matrix $J_{kl}$, so that arrays with holes, edges and position disorder enter only through the couplings. The variance requires a product over the remaining atoms for each of the $N(N-1)$ ordered pairs, so the number of operations grows as $N^3$, which allows exact calculations for arrays of hundreds of atoms.

To find the optimal squeezing, we evaluate $\xi^2$ as a function of time. We locate the optimal time with a geometric scan followed by a golden-section search. We validated the solution against brute-force evolution of the full state for $N\le12$, including arrays with holes and position disorder, and found agreement to machine precision. As a further check, for all-to-all coupling, $J_{kl}=J$, the correlators reduce to the one-axis-twisting expressions~\cite{KitagawaUeda1993}, for instance $\langle\Sx\rangle=\frac N2\cos^{N-1}(2Jt)$, and our implementation reproduces this limit.

\section{Derivation of $F(\kappa,\epsilon)$}
\label{app:derivation}
In this appendix, we derive the formula $F(\kappa,\epsilon)$ from the exact moments of Appendix~\ref{app:validation}. We first write the exact squeezing in units of the coherent state, then expand the cosine products, and finally reduce the remaining sums over atoms and pairs to the two statistics $\kappa$ and $\epsilon$.

\subsection{Exact contrast, twisting correlation and added noise}
Inserting the contrast $C$, the added noise $x$ and the twisting correlation $b$ of Sec.~\ref{sec:formula}, which measure $\langle\Sx\rangle$, $V_{yy}$ and $V_{yz}$ in units of their values in the coherent state, into the minimal variance $V_{\min}$ of Appendix~\ref{app:validation} gives
\begin{equation}
  V_{\min}=\frac N4\Big[1+\frac x2-\sqrt{\frac{x^2}{4}+b^2}\Big],
\end{equation}
and dividing $NV_{\min}$ by $\langle\Sx\rangle^2=N^2C^2/4$ gives Eq.~\eqref{eq:xiCbx}. With the two-spin products
\begin{equation}
  P^\pm_{kl}\equiv\prod_{j\neq k,l}\cos(\theta_{kj}\pm\theta_{lj}),
\end{equation}
the exact moments of Appendix~\ref{app:validation} give the contrast
\begin{equation}
  C=\frac1N\sum_k\prod_{j\neq k}\cos\theta_{kj},\label{eq:Cexact}
\end{equation}
the added noise
\begin{equation}
  x=\frac1{2N}\sum_{k\neq l}\big(P^-_{kl}-P^+_{kl}\big),\label{eq:xexact}
\end{equation}
and the twisting correlation
\begin{equation}
  b=\frac1N\sum_{k\neq l}\sin\theta_{kl}\prod_{j\neq k,l}\cos\theta_{kj}.\label{eq:bexact}
\end{equation}
Up to this point, every expression is exact.

\subsection{Expansion of the cosine products}
Each cosine product is the average of $e^{i\varphi_k}$ over the initial state, where the precession phase $\varphi_k=\sum_j\theta_{kj}s_j$ of atom $k$ is a sum of independent terms with $s_j=\pm1$. Expanding the cosine and then the logarithm gives
\begin{equation}
  \ln\cos\theta=-\frac{\theta^2}{2}-\frac{\theta^4}{12}-\dots .
\end{equation}
The first term sums to the phase variance of atom $k$, which is a collective sum over all partners and is of order one at the optimum. We therefore keep its exponential without expanding it. The second term is small when each atom has several partners and no single bond dominates, and we drop it [approximation (i), small single-bond angles]. Exponentiating the sum of logarithms gives
\begin{equation}
  \prod_{j\neq k}\cos\theta_{kj}\simeq\exp\Big(-\frac12\sum_{j\neq k}\theta_{kj}^2\Big)=e^{-u_k/2},
\end{equation}
with the phase variance of atom $k$
\begin{equation}
  u_k\equiv\sum_{j\neq k}\theta_{kj}^2=4t^2q_k .
\end{equation}
We then replace $q_k$ by its average over the atoms [approximation (ii), the same $q_k$ on every atom], so that $u_k\simeq u\equiv4t^2\qb$.

\subsection{Contrast and twisting correlation}
Inserting the expanded product into Eq.~\eqref{eq:Cexact}, every atom contributes the same factor, and
\begin{equation}
  C\simeq\frac1N\sum_ke^{-u_k/2}\simeq e^{-u/2},
\end{equation}
which is Eq.~\eqref{eq:C}. The product in Eq.~\eqref{eq:bexact} lacks only the factor $\cos\theta_{kl}$, which is close to one for a small bond angle, and $\sin\theta_{kl}\simeq\theta_{kl}$. With $\sum_{l\neq k}\theta_{kl}=2th_k$, this gives
\begin{align}
  b&\simeq e^{-u/2}\,\frac1N\sum_k\sum_{l\neq k}\theta_{kl}\nonumber\\
  &=e^{-u/2}\,2t\,\frac1N\sum_kh_k=2t\,\hb\,e^{-u/2}.
\end{align}
Squaring and multiplying and dividing by $\qb$ gives
\begin{equation}
  b^2=\frac{\hb^{\,2}}{\qb}\,4t^2\qb\,e^{-u}=\kappa\,u\,e^{-u}=\kappa X,
\end{equation}
which is Eq.~\eqref{eq:b}, with $X=u\,e^{-u}$.

\subsection{Added noise}
The same expansion applied to the two-spin products gives
\begin{align}
  \ln P^\pm_{kl}&\simeq-\frac12\sum_{j\neq k,l}\big(\theta_{kj}\pm\theta_{lj}\big)^2\nonumber\\
  &=-\frac12\sum_{j\neq k,l}\big(\theta_{kj}^2+\theta_{lj}^2\big)\mp M_{kl},
\end{align}
where the shared-partner sum
\begin{equation}
  M_{kl}\equiv\sum_{j\neq k,l}\theta_{kj}\theta_{lj}=4t^2\sum_{j\neq k,l}J_{kj}J_{lj}
\end{equation}
measures the partners that atoms $k$ and $l$ share. The first sum equals $u_k+u_l$ minus the single bond $2\theta_{kl}^2$, which we drop under approximation (i). With $u_k\simeq u_l\simeq u$, the two products become $P^\pm_{kl}\simeq e^{-u}e^{\mp M_{kl}}$, and their difference is
\begin{equation}
  P^-_{kl}-P^+_{kl}\simeq2e^{-u}\sinh M_{kl}\simeq2e^{-u}M_{kl}.
\end{equation}
In the last step, we keep only the linear term of $\sinh M_{kl}$, because a single pair shares only a small part of its partners [approximation (iii), small shared-partner terms]. Inserting this into Eq.~\eqref{eq:xexact} gives
\begin{equation}
  x\simeq\frac{e^{-u}}{N}\,4t^2\sum_{k\neq l}\ \sum_{j\neq k,l}J_{kj}J_{lj}.
\end{equation}
The key step is to reorder this triple sum so that the shared partner $j$ is outermost. For fixed $j$, the atoms $k$ and $l$ run over all atoms other than $j$ with $k\neq l$, and since $J$ is symmetric,
\begin{align}
  \sum_{k\neq l}\ \sum_{j\neq k,l}J_{kj}J_{lj}
  &=\sum_j\Big[\Big(\sum_{k\neq j}J_{jk}\Big)^2-\sum_{k\neq j}J_{jk}^2\Big]\nonumber\\
  &=\sum_j\big(h_j^2-q_j\big)=N\big(\overline{h^2}-\qb\big).
\end{align}
This identity holds for any couplings, and it is where the spread of the local fields enters, through $\overline{h^2}=\hb^{\,2}(1+\epsilon)$. Factoring out $u=4t^2\qb$, we obtain
\begin{equation}
  x\simeq u\,e^{-u}\Big[\frac{\hb^{\,2}(1+\epsilon)}{\qb}-1\Big]=X\big[\kappa(1+\epsilon)-1\big],
\end{equation}
which is Eq.~\eqref{eq:x}.

\subsection{The formula and its optimum}
Inserting $C^2=e^{-u}$, $b^2=\kappa X$ and $x=aX$, with $a\equiv\kappa(1+\epsilon)-1$, into Eq.~\eqref{eq:xiCbx} gives
\begin{equation}
  \xi^2(u)=e^{u}g(u),
\end{equation}
where
\begin{equation}
  g(u)=1+\frac{aX}{2}-\sqrt{\frac{a^2X^2}{4}+\kappa X}.
\end{equation}
The array enters only through $\kappa$ and $\epsilon$. Since $u=4t^2\qb$ grows monotonically with time, minimizing over time is the same as minimizing over $u$, which gives Eq.~\eqref{eq:F}. With $dX/du=(1-u)e^{-u}$, the condition $d\xi^2/du=e^u(g+dg/du)=0$ for the optimal phase variance $u^\ast$ reads
\begin{equation}
  g(u^\ast)+(1-u^\ast)e^{-u^\ast}\bigg[\frac a2-\frac{a^2X^\ast/2+\kappa}{2\sqrt{a^2X^{\ast2}/4+\kappa X^\ast}}\bigg]=0,
\end{equation}
with $X^\ast=u^\ast e^{-u^\ast}$. We solve this single equation numerically, and $F=e^{u^\ast}g(u^\ast)$.

The formula also gives the response of the optimal squeezing to a weak spread of the local fields. The spread $\epsilon$ enters only through $x$, and $\xi^2$ is stationary in $u$ at the optimum, so that to first order in $\epsilon$ we obtain $F(\kappa,\epsilon)\simeq F(\kappa,0)+\eta\,\epsilon/C^{\ast2}$. Here $C^{\ast2}=e^{-u^\ast}$ is the squared contrast at the optimal phase variance $u^\ast$ of $F(\kappa,0)$, and the coefficient is
\begin{equation}
  \eta=\frac{\kappa X^\ast}{2}\bigg[1-\frac{(\kappa-1)X^\ast}{\sqrt{(\kappa-1)^2X^{\ast2}+4\kappa X^\ast}}\bigg].
\end{equation}
The coefficient grows with $\kappa$ and approaches one for many effective partners, whereas the contrast at the optimum changes only little with $\kappa$. The slope of $F$ with $\epsilon$ is therefore set mainly by the number of effective partners rather than by the loss of contrast, and it is larger for dipolar couplings than for van der Waals couplings, whose atoms have fewer effective partners.

\subsection{Where the approximations hold}
The derivation uses three approximations, (i) small single-bond angles, so that the $\theta^4$ term of $\ln\cos\theta$ and the single bonds missing from the two-spin products are negligible, (ii) the same $q_k$ on every atom, and (iii) small shared-partner terms, so that $\sinh M_{kl}\simeq M_{kl}$. For couplings that decay with distance, all three hold at the optimum, and the approximate $C$, $b$ and $x$ agree closely with the exact sums of Eqs.~\eqref{eq:Cexact}--\eqref{eq:bexact}. For all-to-all coupling, $J_{kl}=J$, every pair shares all its partners and $M_{kl}=4t^2(N-2)J^2\simeq u$, so approximation (iii) fails. The dropped term $M_{kl}^3/6$ is then the curvature term that limits one-axis twisting~\cite{KitagawaUeda1993}, and $F$ overestimates the squeezing. Since the algebra and the identity for the shared partners are exact, the accuracy of $F$ is set by the dropped terms alone. We therefore regard $F$ as a consistent leading-order expansion whose accuracy is verified against the exact solution, rather than as a controlled expansion in a small parameter.

The single-bond terms and the quartic cumulant are of the same order in the bond angle, but they act on the squeezing in opposite directions. The single-bond terms mainly raise the twisting correlation and thus increase the squeezing, whereas the quartic terms lower the contrast and raise the added noise and thus reduce it. Their effects on $\xi^2$ are of similar size and largely cancel, both for dipolar and for van der Waals couplings and both with and without holes. They must therefore be kept or dropped together, because keeping only one of them leaves an uncompensated correction of the same order, which degrades the prediction rather than improving it.

\subsection{Next order}
The leading order neglects the quartic cumulant of each cosine product, the single bonds that are missing from the products over $j\neq k,l$, and the variation of $q_k$ from atom to atom. Keeping these terms to the next order requires four further statistics of the couplings, the sums $c_j=\sum_iJ_{ij}^3$ and $p_j=\sum_iJ_{ij}^4$, the triangle term $T=N^{-1}\sum_{k,l}J_{kl}^2(J^2)_{kl}$, and the weighted field $(Jh)_k=\sum_jJ_{kj}h_j$. With $s=4t^2$ and $P=s^2\bar p$, the contrast becomes
\begin{equation}
  C^2=e^{-u}\big(1-P/6\big),
\end{equation}
the added noise becomes
\begin{equation}
\begin{split}
  x&=e^{-u}\big(1-\tfrac P6\big)s\Big[\overline{h^2}-\qb+s\big(T+\tfrac23(\overline{ch}-\bar p)\big)\Big]\\
  &\quad-e^{-u}s^2\big[\mathrm{cov}(q,Jh)-\mathrm{Var}\,q\big],
\end{split}
\end{equation}
and the twisting correlation becomes
\begin{equation}
  b=e^{-u/2}\,2t\Big[\big(1-\tfrac{P}{12}\big)\big(\hb+\tfrac{s\,\bar c}{3}\big)-\tfrac s2\,\mathrm{cov}(q,h)\Big],
\end{equation}
where the covariances run over the atoms. Inserted into Eq.~\eqref{eq:xiCbx}, these expressions reduce the typical difference from the exact optimum by about a factor of three across the arrays we tested, which confirms that $F$ is the consistent leading term of this expansion.

\section{Spontaneous-emission correlators}
\label{app:emission}
Spontaneous emission changes the populations of the spins, so, unlike dephasing, it does not simply multiply the closed-system correlators by decay factors. The dynamics nevertheless remains exactly solvable, and in this appendix we derive the correlators that give the exact squeezing in the presence of emission.

We consider the master equation~\eqref{eq:master} with $\gamma=0$, in which each spin decays from $|{\uparrow}\rangle$ to $|{\downarrow}\rangle$ through the jump operator $\sqrt\Gamma\,\sigma^-_k$. Because $H$ is diagonal in the $\sigma^z$ basis, the decay acts on the populations as a classical process, and the populations of different spins decay independently. In the initial state, each spin is found in $|{\uparrow}\rangle$ or $|{\downarrow}\rangle$ with probability $1/2$. A spin in $|{\downarrow}\rangle$ stays there, and a spin in $|{\uparrow}\rangle$ jumps to $|{\downarrow}\rangle$ at a random time $\tau$ with probability density $\Gamma e^{-\Gamma\tau}$. Each spin $j$ thus follows a classical history $z_j(t')=\pm1$, and a partner with coupling $c$ shifts the phase of a coherent spin by $2c\int_0^tz_j(t')\,dt'$, which reduces to $\pm\theta_{kj}$ for $c=J_{kj}$ in the absence of emission. Averaging over the histories of one partner defines the two functions $\phi(c)=\mathbb E\big[e^{2ic\int_0^tz_j\,dt'}\big]$ and $\psi(c)=\mathbb E\big[z_j(t)\,e^{2ic\int_0^tz_j\,dt'}\big]$, which evaluate to
\begin{equation}
  \phi(c)=\tfrac12 e^{-2ict}+\tfrac12\Big[e^{-\Gamma t}e^{2ict}+\Gamma e^{-2ict}\frac{e^{(4ic-\Gamma)t}-1}{4ic-\Gamma}\Big]
\end{equation}
and
\begin{equation}
  \psi(c)=-\tfrac12 e^{-2ict}+\tfrac12\Big[e^{-\Gamma t}e^{2ict}-\Gamma e^{-2ict}\frac{e^{(4ic-\Gamma)t}-1}{4ic-\Gamma}\Big],
\end{equation}
where the first term comes from histories in which the partner starts in $|{\downarrow}\rangle$, the second from histories in which it stays in $|{\uparrow}\rangle$ up to time $t$, and the last from histories in which it decays at a time $\tau<t$, integrated over $\tau$. Both functions also depend on $t$ and $\Gamma$. For $\Gamma\to0$, they reduce to $\phi(c)=\cos2ct$ and $\psi(c)=i\sin2ct$, which for $c=J_{kj}$ are the cosines and sines of the closed system.

A spin whose coherence enters a correlator, such as spin $k$ in $\langle\sigma^+_k\rangle$, must not have decayed, since a jump destroys its coherence, and the anticommutator term of Eq.~\eqref{eq:master} damps this coherence as $e^{-\Gamma t/2}$. In a two-spin coherence, the phases of the bond between $k$ and $l$ cancel as in the closed system, and spin $l$ in $\langle\sigma^+_k\sigma^z_l\rangle$ contributes the function $\psi$ of its own history. Every other spin contributes one factor $\phi$, which gives the single-spin coherence
\begin{equation}
  \langle\sigma^+_k\rangle=\tfrac12e^{-\Gamma t/2}\prod_{j\neq k}\phi(J_{kj}),
\end{equation}
the two-spin coherences
\begin{equation}
  \langle\sigma^+_k\sigma^\pm_l\rangle=\tfrac14e^{-\Gamma t}\prod_{j\neq k,l}\phi(J_{kj}\pm J_{lj}),
\end{equation}
and the correlation of a coherence with a population
\begin{equation}
  \langle\sigma^+_k\sigma^z_l\rangle=\tfrac12e^{-\Gamma t/2}\,\psi(J_{kl})\prod_{j\neq k,l}\phi(J_{kj}).
\end{equation}
The populations follow from the decay alone,
\begin{equation}
  \langle\sigma^z_k\rangle=e^{-\Gamma t}-1,
\end{equation}
and
\begin{equation}
  \langle\sigma^z_k\sigma^z_l\rangle=\langle\sigma^z\rangle^2 .
\end{equation}
The decay moves spins only downward, so the mean population $\langle\sigma^z_k\rangle$ decreases from zero toward $-1$, and $\langle\sigma^z_k\sigma^z_l\rangle$ factorizes because different spins decay independently. Because the decay biases the partners toward $|{\downarrow}\rangle$, their phase shifts no longer average to zero, $\phi$ acquires an imaginary part, and the mean spin rotates in the $xy$ plane while it tilts toward $-z$. We therefore compute the squeezing from the full covariance matrix of $\mathbf S$ in the plane perpendicular to the mean spin. This solution is of the type derived by Foss-Feig \textit{et al.}~\cite{FossFeig2013}, and we include it for completeness. We checked it against a direct numerical integration of the master equation~\eqref{eq:master} for small arrays and found agreement to machine precision.

The formula with spontaneous emission in Sec.~\ref{sec:decoherence} uses three statistics of the time integral $Z$ of the $z$ component of a single decaying spin. Averaging over the same histories as above, we obtain its mean
\begin{equation}
  m_1=\mathbb E[Z]=\frac{1-e^{-\Gamma t}}{\Gamma}-t,
\end{equation}
its variance
\begin{equation}
  v=\mathrm{Var}\,Z=4\Big(\frac{1-e^{-\Gamma t}(1+\Gamma t)}{\Gamma^2}-\frac{(1-e^{-\Gamma t})^2}{4\Gamma^2}\Big),
\end{equation}
and its covariance with the final $z$ component
\begin{equation}
  w=\mathrm{Cov}\big(z(t),Z\big)=2e^{-\Gamma t}\Big[t-\frac{1-e^{-\Gamma t}}{2\Gamma}\Big].
\end{equation}
The mean $z$ component of a decaying spin, $\bar z=e^{-\Gamma t}-1$, equals the single-spin population $\langle\sigma^z_k\rangle$ given above.


\begin{thebibliography}{99}
\bibitem{Wineland1994} D.~J.~Wineland, J.~J.~Bollinger, F.~L.~Moore, and D.~J.~Heinzen, Phys. Rev. A \textbf{50}, 67 (1994).
\bibitem{KitagawaUeda1993} M.~Kitagawa and M.~Ueda, Phys. Rev. A \textbf{47}, 5138 (1993).
\bibitem{Gil2014} L.~I.~R.~Gil, R.~Mukherjee, E.~M.~Bridge, M.~P.~A.~Jones, and T.~Pohl, Phys. Rev. Lett. \textbf{112}, 103601 (2014).
\bibitem{Hines2023} J.~A.~Hines \textit{et al.}, Phys. Rev. Lett. \textbf{131}, 063401 (2023).
\bibitem{Eckner2023} W.~J.~Eckner \textit{et al.}, Nature \textbf{621}, 734 (2023).
\bibitem{Bornet2023} G.~Bornet \textit{et al.}, Nature \textbf{621}, 728 (2023).
\bibitem{Bilitewski2021} T.~Bilitewski \textit{et al.}, Phys. Rev. Lett. \textbf{126}, 113401 (2021).
\bibitem{FossFeig2013} M.~Foss-Feig, K.~R.~A.~Hazzard, J.~J.~Bollinger, and A.~M.~Rey, Phys. Rev. A \textbf{87}, 042101 (2013).
\bibitem{vandenWorm2013} M.~van den Worm, B.~C.~Sawyer, J.~J.~Bollinger, and M.~Kastner, New J. Phys. \textbf{15}, 083007 (2013).
\bibitem{FossFeig2016} M.~Foss-Feig, Z.-X.~Gong, A.~V.~Gorshkov, and C.~W.~Clark, arXiv:1612.07805 (2016).
\bibitem{Perlin2020} M.~A.~Perlin, C.~Qu, and A.~M.~Rey, Phys. Rev. Lett. \textbf{125}, 223401 (2020).
\bibitem{Young2023} J.~T.~Young, S.~R.~Muleady, M.~A.~Perlin, A.~M.~Kaufman, and A.~M.~Rey, Phys. Rev. Research \textbf{5}, L012033 (2023).
\bibitem{Schlosser2001} N.~Schlosser, G.~Reymond, I.~Protsenko, and P.~Grangier, Nature \textbf{411}, 1024 (2001).
\bibitem{Barredo2016} D.~Barredo, S.~de~L\'es\'eleuc, V.~Lienhard, T.~Lahaye, and A.~Browaeys, Science \textbf{354}, 1021 (2016).
\bibitem{Endres2016} M.~Endres \textit{et al.}, Science \textbf{354}, 1024 (2016).
\bibitem{VanDamme2021} J.~Van Damme, X.~Zheng, M.~Saffman, M.~G.~Vavilov, and S.~Kolkowitz, Phys. Rev. A \textbf{103}, 023106 (2021).
\bibitem{Begg2026} S.~E.~Begg, B.~K.~Ghosh, C.~Zu, C.~Zhang, and M.~Kolodrubetz, arXiv:2601.10703 (2026).
\bibitem{Solfanelli2026} A.~Solfanelli, A.~Smerzi, P.~Zoller, and N.~Defenu, arXiv:2605.03032 (2026).
\bibitem{Zeiher2016} J.~Zeiher \textit{et al.}, Nat. Phys. \textbf{12}, 1095 (2016).
\bibitem{Borish2020} V.~Borish, O.~Markovi\'c, J.~A.~Hines, S.~V.~Rajagopal, and M.~Schleier-Smith, Phys. Rev. Lett. \textbf{124}, 063601 (2020).
\bibitem{Britton2012} J.~W.~Britton \textit{et al.}, Nature \textbf{484}, 489 (2012).
\bibitem{Bohnet2016} J.~G.~Bohnet \textit{et al.}, Science \textbf{352}, 1297 (2016).
\bibitem{Henkel2010} N.~Henkel, R.~Nath, and T.~Pohl, Phys. Rev. Lett. \textbf{104}, 195302 (2010).
\bibitem{Pupillo2010} G.~Pupillo, A.~Micheli, M.~Boninsegni, I.~Lesanovsky, and P.~Zoller, Phys. Rev. Lett. \textbf{104}, 223002 (2010).
\bibitem{Goldschmidt2016} E.~A.~Goldschmidt \textit{et al.}, Phys. Rev. Lett. \textbf{116}, 113001 (2016).
\bibitem{Jaksch2000} D.~Jaksch \textit{et al.}, Phys. Rev. Lett. \textbf{85}, 2208 (2000).
\bibitem{Lukin2001} M.~D.~Lukin \textit{et al.}, Phys. Rev. Lett. \textbf{87}, 037901 (2001).
\bibitem{JohnsonRolston2010} J.~E.~Johnson and S.~L.~Rolston, Phys. Rev. A \textbf{82}, 033412 (2010).
\bibitem{Honer2010} J.~Honer, H.~Weimer, T.~Pfau, and H.~P.~B\"uchler, Phys. Rev. Lett. \textbf{105}, 160404 (2010).
\bibitem{Saffman2010} M.~Saffman, T.~G.~Walker, and K.~M{\o}lmer, Rev. Mod. Phys. \textbf{82}, 2313 (2010).
\bibitem{Schachenmayer2015} J.~Schachenmayer, A.~Pikovski, and A.~M.~Rey, Phys. Rev. X \textbf{5}, 011022 (2015).
\bibitem{Truong2026} T.-V.~Truong and H.-L.~Nguyen, arXiv:2611.34004 (2026).
\bibitem{CodeRepo} H.-L.~Nguyen and T.-V.~Truong, \href{https://github.com/Identity-Operator/ising-squeezing-arrays}{github.com/Identity-Operator/ising-squeezing-arrays} (2026).
\bibitem{CodeZenodo} H.-L.~Nguyen and T.-V.~Truong, Zenodo (2026), \href{https://doi.org/10.5281/zenodo.24652761}{doi:10.5281/zenodo.24652761}.
\end{thebibliography}
\end{document}